\documentclass[aps,prb,twocolumn,superscriptaddress,showpacs]{revtex4}

\usepackage{graphicx}

\begin{document}


\title {Spin Dynamics in the Magnetoelectric Effect LiCoPO$_4$ Compound}



\author{Wei Tian}
\affiliation{Ames Laboratory and Department of Physics and
Astronomy, Iowa State University, Ames, Iowa 50011, USA}

\author{Jiying Li}
\affiliation{NCNR, National Institute of Standards and Technology,
Gaithersburg, MD 20899, USA} \affiliation{Dept. of Materials Science
and Engineering, University of Maryland, College Park, MD 20742,
USA}

\author{Jeffrey W. Lynn}
\affiliation{NCNR, National Institute of Standards and Technology,
Gaithersburg, MD 20899, USA}

\author{Jerel L. Zarestky}
\affiliation{Ames Laboratory and Department of Physics and
Astronomy, Iowa State University, Ames, Iowa 50011, USA}

\author{David Vaknin}
\affiliation{Ames Laboratory and Department of Physics and
Astronomy, Iowa State University, Ames, Iowa 50011, USA}

\begin{abstract}
Inelastic neutron scattering (INS) experiments were performed to
investigate the spin dynamics in magnetoelectric effect (ME)
LiCoPO$_4$ single crystals. Weak dispersion was detected in the
magnetic excitation spectra along the three principal
crystallographic axes measured around the (0 1 0) magnetic
reflection. Analysis of the data using linear spin-wave theory
indicate that single-ion anisotropy in LiCoPO$_4$ is as important as
the strongest nearest-neighbor exchange coupling. Our results
suggest that Co$^{2+}$ single-ion anisotropy plays an important role
in the spin dynamics of LiCoPO$_4$ and must be taken into account in
understanding its physical properties. High resolution INS
measurements reveal an anomalous low energy excitation that we
hypothesize may be related to the magnetoelectric effect of
LiCoPO$_4$.
\end{abstract}

\maketitle
\section{Introduction}
LiCoPO$_4$ is an antiferromagnetic (AFM) insulator belonging to the
olivine family of lithium orthophosphates that share the general
chemical formula Li{\it M}PO$_4$ ({\it M} = Mn$^{2+}$, Fe$^{2+}$,
Co$^{2+}$, Ni$^{2+}$) with four formula units per unit cell
\cite{Megaw-1973,Santoro-1966-1967}. These materials continue to
attract much attention due to their exceptionally large
magnetoelectric (ME) effect and the anomalies exhibited in the ME
coefficients as a function of temperature and magnetic
field\cite{Mercier-1967,Mercier-1971,Rivera-1994}. To date, it
remained an open question whether the anomalies observed in the ME
effect of Li{\it M}PO$_4$ are intrinsic due to the particular local
environment surrounding the transition metal ions in
Li\textit{M}PO$_4$ or due to domain formation structure. The local
environment can be slightly distorted in a magnetic (electric) field
by virtue of the spin-orbit coupling, giving rise to a collective
ferroelectric response (magneto-ferro-elastic
effect)\cite{Rado,Jensen2007}. The recent domain structure observed
by second harmonic generation (SHG) in LiCoPO$_4$ was attributed to
coexisting AFM and ferrotoroidic domains which may play a role in
giving rise to the ME effect\cite{Fiebig2007}. In addition to the
strong ME effect, this family of materials also exhibits intriguing
magnetic properties. At low temperatures, Li{\it M}PO$_4$ systems
undergo transitions to AFM long-range-order (LRO), adopting similar
magnetic structures, differing only in spin orientation. For
example, with increasing temperature, LiNiPO$_4$ first undergoes a
first-order commensurate-incommensurate (C-IC) phase transition at
$T_N \approx 20.8$ K changing from a co-linear AFM state to a
long-range IC order state, followed by a second-order phase
transition from long-range IC to short-range IC order at $T_{IC}
\approx$ 21.7 K \cite{Vaknin-PRL-2004}. On the other hand,
LiMnPO$_4$ undergoes an AFM LRO transition at $T_N \approx$ 34 K as
well as a field induced spin-flop transition
\cite{Jiying-Mntobepublish}. Weak ferromagnetism \cite{Kharcheno},
an ME ``butterfly loop" anomaly \cite{Kornev-2000}, and strong
magnetic anisotropy have been observed in LiCoPO$_4$
\cite{Rivera-1998} which also exhibits the largest ME coefficient
\cite{Rivera-1994} among its counterpart compounds, making it of
particular interest to study.

\begin{figure}
\centering\includegraphics[width=2.0in]{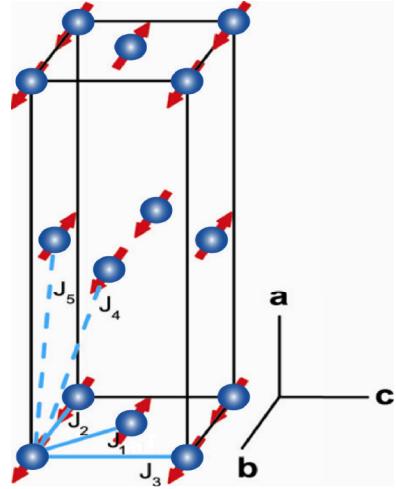}
\caption{\label{fig:magcell}(Color online) Magnetic unit cell of
LiCoPO$_4$ displaced (0.25 0.25 0) r.l.u compared to the atomic unit
cell. Only the Co$^{2+}$ magnetic ions are shown for clarity. The
intra-plane (solid) and inter-plane (dashed) magnetic exchange
interactions considered in the spin Hamiltonian
(Eq.~\ref{eq:spinhamiltonian}) are labeled. It is assumed that the
spin is oriented strictly along the crystallographic \textit{b}-axis
in the model calculation.}
\end{figure}

LiCoPO$_4$ crystallizes in an orthorhombic symmetry, space group
{\it Pnma} (no. 62) at room temperature, with lattice parameters $a$
= 10.093, $b$ = 5.89, and $c$ = 4.705 ${\AA}$. The structure
consists of buckled CoO layers stacked along the crystallographic
\textit{a}-axis and the magnetic Co$^{2+}$ ($S$ = 3/2) ions are
surrounded by oxygen ions in a strongly distorted CoO$_6$ octahedral
coordination. LiCoPO$_4$ develops AFM LRO at $T_N \approx 21.8$ K
\cite{Kornev-2000,Vaknin-PRB-2002}. Earlier studies have indicated a
simple two-sublattice AFM state below $T_N$ with spins aligned along
the $b$-axis (whereas LiFePO$_4$, LiMnPO$_4$, and LiNiPO$_4$ have
spins oriented along the $b$, $a$, and $c$-axis
respectively)\cite{Santoro-1966-1967,Vaknin1999}. Fig.\
\ref{fig:magcell} illustrates the magnetic structure of LiCoPO$_4$.
For simplicity, only the Co$^{2+}$ magnetic ions are shown. As
depicted in Fig.~\ref{fig:magcell}, different magnetic exchange
pathways are at play in this compound. Within the buckled CoO layer,
nearest-neighbor Co$^{2+}$ ions are strongly coupled ($J_1$) through
Co-O-Co super-exchange interactions, while additional inplane
magnetic interactions between next-nearest-neighbors ($J_2$, $J_3$)
are mediated through the $PO_4$ group. Between adjacent layers, the
magnetic couplings of nearest-neighbor ions ($J_4$, $J_5$) are also
mediated via the $PO_4$ groups. The $PO_4$ coupling are found to be
rather strong and cannot be neglected
\cite{Zarestky-2001,Vaknin-PRL-2004}. Due to its layered structure,
LiCoPO$_4$ exhibits properties between two- and three-dimensional
($2D$ and $3D$) magnetic systems.

Although the physical properties of LiCoPO$_4$ have been studied
extensively in the past, there remain a number of puzzles in this
compound \cite{Wiegelmann-thesis}. Recent magnetoelectric
\cite{Rivera-1994,Kornev-2000}, magneto-optic
\cite{Kharchenko-2000}, and magnetic property \cite{Kharchenko-2001}
studies do not agree with the originally proposed co-linear AFM
structure and suggest a more complex magnetic structure for
LiCoPO$_4$. Neutron diffraction studies suggest the moments in
LiCoPO$_4$ in the AFM phase are not strictly aligned along the
$\textit{b}$-axis but are uniformly rotated from this axis by a
small angle ($\sim$ 4.6$^\circ$) \cite{Vaknin-PRB-2002}. The
observation of weak ferromagnetism and an ME ``butterfly loop"
anomaly further motivated studies of LiCoPO$_4$ both experimentally
and theoretically \cite{Gufan-1987,Kornev-1999,Kornev2-1999}. In
this paper, we report single crystal inelastic neutron scattering
(INS) studies that yield the microscopic magnetic interactions in
LiCoPO$_4$. The data were analyzed within the linear spin wave
approximation using a spin Hamiltonian explicitly including the
intra-, inter-plane nearest neighbor, next-nearest-neighbor exchange
interactions and single-ion anisotropy, which are determined in this
study.

\section{EXPERIMENTAL TECHNIQUES}

All measurements reported here were carried out on LiCoPO$_4$ single
crystals. Large crystals were grown for INS experiments by a LiCl
flux method similar to that reported in Ref.
\onlinecite{Fomin-2002}. High purity starting materials of CoCl$_2$
(99.999$\%$, MV laboratory), Li$_3$PO$_4$ (99.999$\%$ Aldrich) and
LiCl (99.999$\%$) were thoroughly ground together at a molar ratio
of Li$_3$PO$_4$:CoCl$_2$:LiCl = 1:1:1 and sealed in a Pt crucible
under Ar atmosphere. Note that the reaction of Li$_3$PO$_4$ +
CoCl$_2$ + LiCl $\longrightarrow$ LiCoPO$_4$ + 3LiCl yields a molar
ratio of 1:3 between LiCoPO$_4$ and the flux material LiCl which we
found crucial in growing large LiCoPO$_4$ single crystals. The mixed
powder is pre-melted at 800$^{\circ}$C and slowly heated to
900$^{\circ}$C. The crucible was maintained at 900$^{\circ}$C for 10
hours and then slowly cooled to 640$^{\circ}$C at a cooling rate of
0.7$^{\circ}$C/hr and then furnace-cooled to room temperature. Large
purple color LiCoPO$_4$ single crystals were obtained and extracted
by dissolving LiCl in water. The crystals were characterized by
X-ray diffraction measurements and found to be of pure single phase.
The lattice parameters determined from our neutron diffraction
measurements $a$ = 10.159, $b$ = 5.9, and $c$ = 4.70 ${\AA}$ at 8 K
are in good agreement with prior results\cite{Vaknin-PRB-2002}.

Two LiCoPO$_4$ single crystals grown from the same batch were used
for the INS experiments. Sample $\#$1, $m \approx 0.8$ g, and sample
$\#$2, $m \approx 0.4$ g, were oriented in the ($H$ $K$ 0) and (0
$K$ $L$) scattering plane, respectively. INS experiments were
performed using the HB1A triple-axis spectrometer (TAS) at the High
Flux Isotope Reactor (HFIR) neutron scattering facility at Oak Ridge
National Laboratory, the BT7 thermal TAS, and the SPINS cold TAS at
the NIST Center for Neutron Research at the National Institute of
Standards Technology (NCNR). The magnetic excitations along the ($H$
1 0) and (0 $K$ 0) directions were measured using the HB1A TAS on
sample $\#$1 which was mounted on a thin aluminum disk, sealed in an
aluminum sample can and kept under helium atmosphere and cooled
using a closed-cycle He refrigerator. Collimations of
$48'$-$48'$-$sample$-$40'$-$68'$ downstream from reactor to detector
were used throughout the experiment. Constant wave-vector scans were
performed at $T$ = 8 K (fully ordered state) and $T$ = 35 K (well
above T$_N$). Spin excitations along the (0 1 $L$) direction were
measured using the BT7 TAS at NIST on sample $\#$2. A fixed final
energy of $E_f$ = 14.7 meV and a $open$-$50'$-$sample$-$50'$-$open$
collimation were used with pyrolytic graphite (PG(0 0 2)) analyzer
crystals in flat mode. High resolution INS measurements were carried
out on sample $\#$1 using the SPINS TAS with a fixed final energy,
$E_f$ = 5 meV. A collimation of $open$-$80'$-$sample$-$80'$-$open$
was used with a cold Be filter in the scattering beam. Polarized
neutron scattering experiments were also performed using the BT7 TAS
with a fixed final energy of $E_f$ = 13.7 meV to clarify the nature
of the $\sim$ 1.2 meV low energy excitation. The polarization
analysis technique as applied to this study is discussed in Ref.
\onlinecite{polarize-technique1,polarize-technique2,polarize-technique3}.
$^3$$He$ spin filters (polarizer) are mounted before and after the
sample with a spin flipper in the incident beam. The sample is
maintained in a horizontal or vertical magnetic guide field of
$\sim$5 Oe such that the neutron polarization $\bf{\hat{p}}$ is
parallel to the momentum transfer $\bf{Q}$, $\bf{\hat{p}}
\parallel \bf{Q}$ when a horizontal field is applied at the sample
position, or $\bf{\hat{p}} \bot \bf{Q}$ when a vertical magnetic
field is applied. With the spin flipper off, we measure the (++)
non-spin-flip scattering. On the other hand, with the spin flipper
on, the (-+) spin-flip scattering is measured. Sample $\#$2 oriented
in the (0 $K$ $L$) scattering plane and a
$open$-$50'$-$sample$-$80'$-$open$ collimation were used throughout
the polarization measurements. Constant wave-vector scans were
carried out with both (++) and (-+) configurations. All measurement
results have been normalized to a beam monitor count.

\section{EXPERIMENTAL RESULTS and MODELING}

\begin{figure} [!ht]
\centering\includegraphics[width=3.1in]{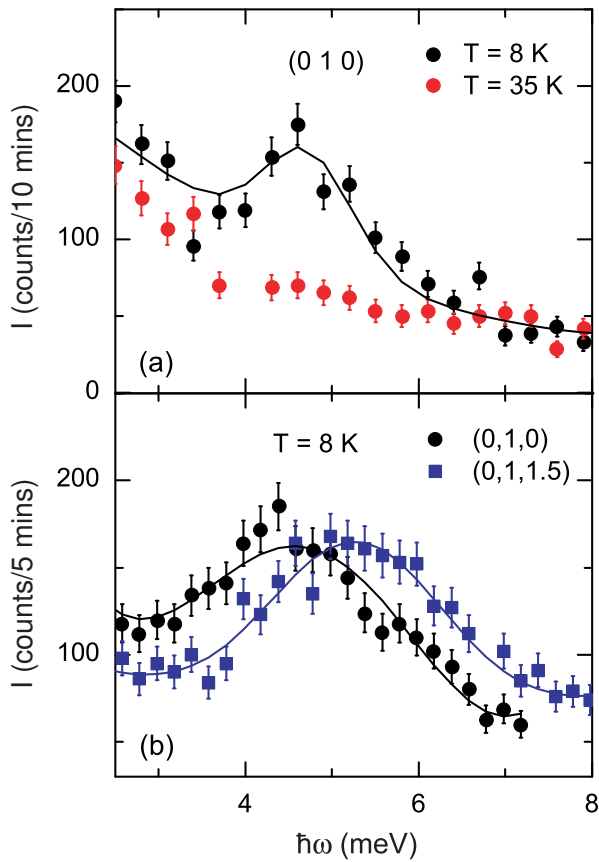}
\caption{\label{fig:spinwave}(Color online) Representative constant
wave-vector scans plotted as scattering intensity versus energy
transfer between $\hbar\omega$ = 2.5 meV and 8 meV. (a). Temperature
dependence of the $\sim$ 4.7 meV excitation measured at (0 1 0) at
$T$ = 8 K and 35 K using the HB1A TAS. (b). $T$ = 8 K constant
wave-vector scans using the BT7 TAS measured at (0 1 0) and (0 1
1.5) showing a single magnetic excitation that exhibits modest but
unambiguous dispersion along the (0 1 $L$) direction. Intensities
were normalized to the incident neutron flux by counting against
neutron monitor counts.}
\end{figure}

Representative constant wave-vector scans with energy transfer
between $\hbar\omega$ = 2.5 meV and 8 meV measured using the HB1A
and BT7 TAS are shown in Fig.~\ref{fig:spinwave}. Note that error
bars in this paper are statistical in origin and represent one
standard deviation. Fig.~\ref{fig:spinwave} (a) shows the
temperature dependence of the magnetic excitation measured at (0 1
0). At $T$ = 8 K in the fully ordered phase, a single excitation
with energy transfer of $\hbar\omega$ $\approx$ 4.7 meV is detected.
At a temperature well above $T_N$ ($T$ = 35 K) the peak disappears
demonstrating the excitation is magnetic in origin. Below $T_N$,
further detailed measurements at (0 1 0) as a function of
temperature indicate no significant temperature dependence of this
excitation. Fig.~\ref{fig:spinwave} (b) shows the $T$ = 8 K constant
wave-vector scans measured at (0 1 0) and (0 1 1.5), which typically
correspond to the minimum and maximum spin wave excitations. The
data clearly show that the excitation observed at $\hbar\omega$
$\approx$ 4.7 meV at (0 1 0) shifts to higher energy transfer
$\hbar\omega$ $\approx$ 5.3 meV at (0 1 1.5), yielding a maximum
energy shift of $\sim$ 0.6 meV. This indicates that the overall
dispersion along the (0 1 $L$) direction is modest compared to an
exchange energy of $kT$$_N$ $\sim$ 2 meV. Similar behaviors were
observed for the dispersions along the (0 $K$ 0) and ($H$ 1 0)
directions. Note that data in Fig.~\ref{fig:spinwave} (a) shows
strong scattering at 2.5 meV, the lowest data point plotted. We will
show later that it is from scattering of a low energy excitation at
$\sim$ 1.2 meV which has been observed in recent SPINS high
resolution measurements.

\begin{figure*}[tbp!]
\centering\includegraphics[width=6.6in]{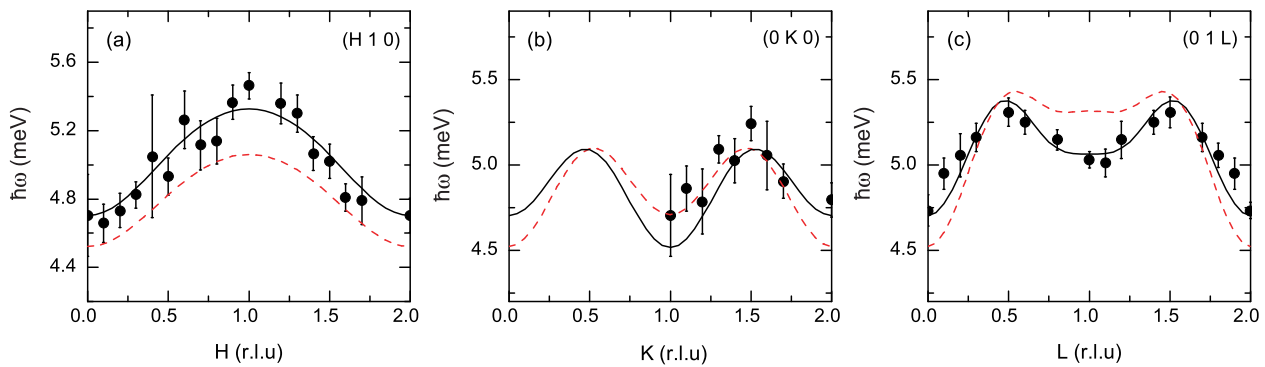}
\caption{\label{fig:disp}(Color online) Spin-wave dispersion curves
along three reciprocal directions constructed from a series constant
wave-vector scans measured at $T$ = 8 K. Data points are obtained
based upon a Gaussian peak approximation. The solid and dash lines
are calculations based upon a global fit to the linear spin wave
approximation theory as described in the text.}
\end{figure*}

To determine the magnetic excitation spectra along all three
principal axes directions, a series of constant wave-vector scans
were carried out in the ($H$ $K$ 0) and (0 $K$ $L$) scattering
planes at $T$ = 8 K around the (0 1 0) magnetic reflection. Fig.
\ref{fig:disp} depicts the ground state magnetic dispersion
relations along the ($H$ 1 0), (0 $K$ 0), and (0 1 $L$) directions
constructed from energy scans at constant wave-vector. We determined
the peak positions assuming Gaussian peak-shapes that were fit to
each of the constant wave-vector scans measured. For both the HB1A
and BT7 triple-axis-spectrometers, the energy resolution at the
elastic position was $\Delta E \approx$ 1 meV. The experimental
uncertainties had a significant effect in the theoretical modeling
as described in the text below. The measured spectra indicate a spin
wave excitation of $\hbar\omega$ $\approx$ 4.7 $\pm 0.24$ meV at (0
1 0) which vanishes abruptly above $T_N$, while modest dispersion
was observed along all three principal symmetry directions, with the
scale in Fig.~\ref{fig:disp} being chosen to best exhibit the
dispersion that falls within a band of 0.8 meV. This relatively weak
dispersion suggests an Ising-like model in LiCoPO$_4$; in a pure
Ising model the magnetic excitations are completely dispersionless.

To analyze the measured spin wave dispersion curves of LiCoPO$_4$
using linear spin wave theory, we consider the different magnetic
exchange interactions as illustrated in Fig.~\ref{fig:magcell} and
assume an AFM ground state with spins pointing strictly along the
$b$-axis. Taking into account the intra-plane and inter-plane
nearest-neighbor, next-nearest-neighbor interactions, and the
single-ion anisotropy, the Spin Hamiltonian can be expressed in the
following form:
\begin{equation}
{\mathcal H} = \sum_{i,j} J_{ij}{\mathbf S}_{i}\cdot {\mathbf S}_{j}
+ \sum_{i,\alpha} D_{\alpha}(S_{i}^{\alpha})^2,
\label{eq:spinhamiltonian}
\end{equation}
where $D_{\alpha}$ ($\alpha = x, y, z$) represents the single-ion
anisotropy along the $x$ , $y$, and $z$-directions. In order to have
the spins pointing along the $z$-axis in the model calculation, the
Cartesian coordinate $x$, $y$, and $z$ directions are defined to be
along the crystallographic $a$, $c$, and $b$ directions
respectively. The zero point of the energy spectrum is chosen such
that $D_z$=0. Within a linear spin wave approximation, the derived
spin wave dispersion from Eq.~\ref{eq:spinhamiltonian} is given by:
\begin{equation}
\hbar\omega = \sqrt{A^2-(B\pm C)^2}. \label{eq:dispersion}
\end{equation}
where\\
 A = $4S(J_1+J_5)-2S[J_2(1-cos(\textbf{q}\cdot
\textbf{r}_5))+J_3(1-cos(\textbf{q}\cdot
\textbf{r}_6))+J_4(2-cos(\textbf{q}\cdot
\textbf{r}_7)-cos(\textbf{q}\cdot \textbf{r}_8))]+D_xS+D_yS$,\\
B = $D_xS-D_yS$,\\
C = $2J_1S(cos(\textbf{q}\cdot \textbf{r}_1)+cos(\textbf{q}\cdot
\textbf{r}_2))+2J_5S(cos(\textbf{q}\cdot
\textbf{r}_3)+cos(\textbf{q}\cdot \textbf{r}_4))$.\\
and $\textbf{r$_i$}$ denotes the vectors directed between two Co$^{2+}$ ions\\
$\textbf{r}_1=(0, b/2, c/2)$, $\textbf{r}_2=(0, b/2, -c/2)$,\\
$\textbf{r}_3=(a/2, b/2, 0)$, $\textbf{r}_4=(a/2, -b/2, 0)$,\\
$\textbf{r}_5=(0, b, 0)$, $\textbf{r}_6=(0, 0, c)$,\\
$\textbf{r}_7=(a/2, 0, c/2)$, $\textbf{r}_8=(a/2, 0,-c/2)$.

Non-linear-least-squares fits of the spin-wave dispersion expressed
by Eq.~\ref{eq:dispersion} to the observed magnetic spectra yields:
$J_1=0.743\pm0.187$ meV, $J_2 = 0.105 \pm$ 0.159 meV, $J_3 = 0.194
\pm$ 0.131 meV, $J_4 = -0.163 \pm$ 0.08 meV, $J_5 = -0.181 \pm$
0.125 meV, $D_x = 0.718 \pm$ 0.192 meV, $D_y = 0.802 \pm$ 0.208 meV.
The obtained microscopic interaction parameter $J_1$ is
significantly larger than $J_4$ and $J_5$ consistent with previous
observations that the magnetic behavior of LiCoPO$_4$ is
intermediate between a $2D$ and $3D$ system. Moreover, the obtained
positive value of $J_1$ indicates in-plane AFM nearest-neighbor
coupling, whereas negative $J_4$ and $J_5$ values suggest
inter-plane FM coupling along the $a$-axis consistent with the
magnetic structure, where the magnetic unit cell is doubled along
the $b$- and $c$-axes, but not along the $a$-axis. $J_2$ and $J_3$
have the same sign as $J_1$ indicating that they compete with $J_1$
and may cause frustration, however, they are relatively weak
compared to $J_1$ ($J_2/J_1 \approx 0.14$, $J_3/J_1 \approx 0.26$).
Both $D_x$ and $D_y$ are positive favoring a ground state with the
magnetic moment along the $b$-axis consistent with the elastic
magnetic neutron scattering results. An important result in our
study is the large values of the single-ion anisotropy compared to
the nearest-neighbor coupling $D_x \sim D_y \sim J_1$. Although
strong single-ion anisotropy in LiCoPO$_4$ has been suggested by
several models \cite{Rado,Bichurin}, this study provides
experimental evidence that the single-ion anisotropy is as important
as the strongest magnetic exchange interaction in LiCoPO$_4$. Such
relatively strong anisotropy may split the $S=3/2$ quartet of the
Co$^{2+}$ ion into two doublets rendering the suggested Ising-like
character to LiCoPO$_4$ \cite{Vaknin-PRB-2002}.

\begin{figure} [!ht]
\centering\includegraphics[width=3.0in]{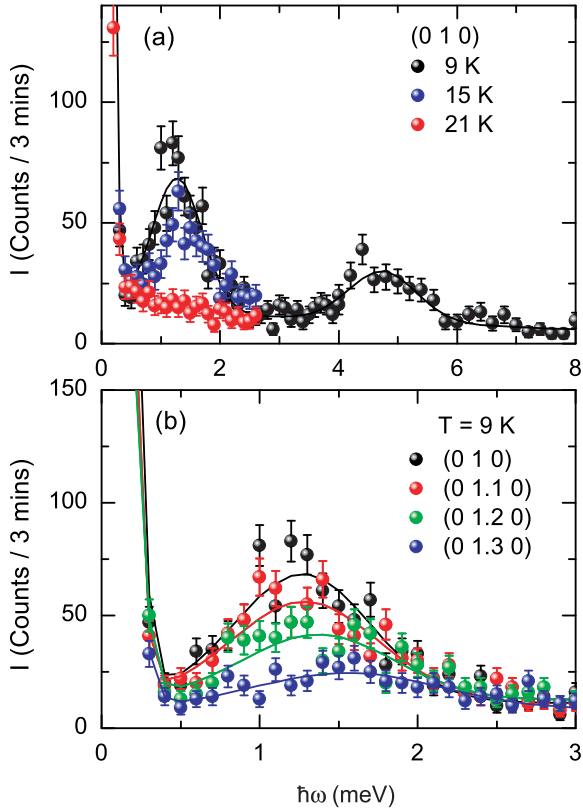}
\caption{\label{fig:SPINS-data}(Color online) SPINS high resolution
measurements of LiCoPO$_4$. (a) Constant wave-vector scans measured
at (0 1 0) at $T$ = 9 K, 15 K, and 21 K indicating a second low
energy excitation at $\hbar\omega$ $\sim$ 1.2 meV below $T_N$. (b)
Weak dispersion observed in the constant wave-vector scans measured
along the (0 $K$ 0) direction at $T$ = 9 K.}
\end{figure}

The ``$\pm$" sign in Eq.~(\ref{eq:dispersion}) ($B\pm C$) indicates
that there are two non-degenerate spin wave branches which come
directly from the different values of $D_x$ and $D_y$. Using the
obtained best-fit parameters, the calculated dispersion curves of
the two branches are plotted as solid lines (``$B-C$" branch) and
dash lines (``$B+C$" branch ) in Fig.~\ref{fig:disp}. The two
calculated spin wave branches predict a maximum separation of $\sim$
0.3 meV at (0 1 0), (1 1 0), and (0 1 1). In the thermal neutron TAS
measurements using HB1A and BT7 with a resolution of $\sim$ 1 meV,
only one excitation was observed at these wave vectors. We have two
high resolution measurements using SPINS TAS, fixed E$_f$ = 5 meV
with a resolution of $\sim$ 0.28 meV, at (1 1 0) and (0 1 0) with
energy transfer up to 8 meV. The constant wave vector scan at (0 1
0) at $T$ = 9 K is shown in Fig.~\ref{fig:SPINS-data} (a). At (0 1
0), where the model predicts the maximum separation between these
two branches, only one excitation around $\sim$ 4.7 meV is observed.
The additional low energy excitation observed at $\sim$ 1.2 meV does
not agree with the model and is discussed below. Our results could
not resolve the two branches for two possible reasons. First, the
second excitation may be very weak in intensity, and our model does
not predict intensities. Second, the intrinsic linewidth of the
observed excitations are broader than the resolution ($\sim$ 1 meV)
suggestive of contributions from both branches overlap and cause the
broadening.

\begin{figure} [!ht]
\centering\includegraphics[width=3.0in]{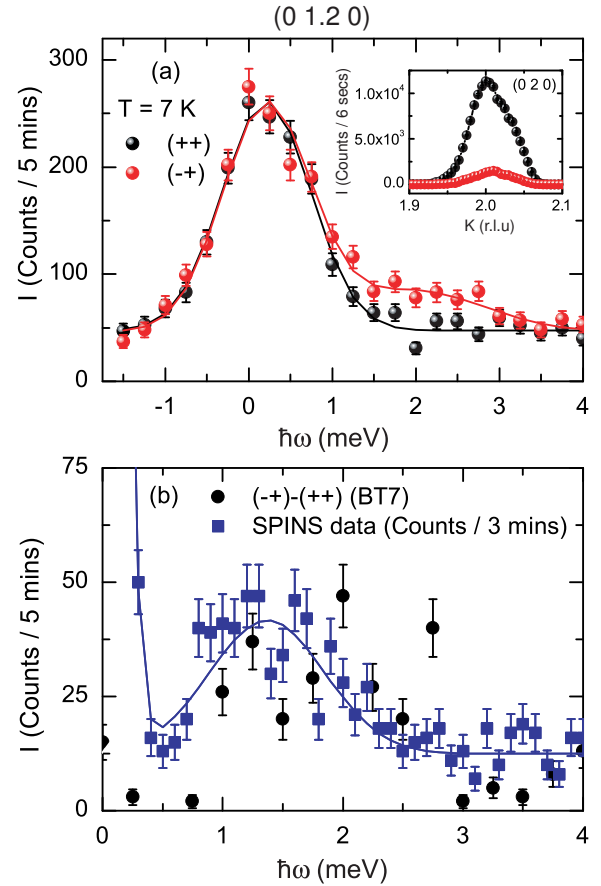}
\caption{\label{fig:polarize}(Color online) BT7 polarized neutron
data measured at (0 1.2 0) in the horizontal field configuration,
$\bf{\hat{p}}\parallel \bf{Q}$. (a) Comparison of spin-flip (-+) and
non-spin-flip (++) scattering measured at $T$ = 7 K indicating the
magnetic origin of the second low energy excitation. Inset:
Spin-flip (-+) and non-spin-flip (++) scattering of the nuclear
reflection (0 2 0). (b) Comparison of SPINS high resolution data and
the BT7 subtracted polarized data, the non-spin-flip (++) data was
subtracted from the spin-flip (-+) data.}
\end{figure}

High resolution measurements on SPINS show an anomalous low energy
excitation below $T_N$ that does not fit in the linear spin wave
model. Fig.\ \ref{fig:SPINS-data} (a) depicts the $T$ = 9 K, 15 K,
and 21 K data measured at (0 1 0). In addition to the $\hbar\omega$
$\approx$ 4.7 meV excitation, the $T$ = 9 K SPINS data clearly show
a low energy excitation at $\hbar\omega$ $\approx$ 1.2 meV. The peak
position of this excitation is practically temperature independent
but the peak intensity decreases with increasing temperature and
eventually the peak vanishes above $T_N$. It also shows weak
dispersion along all three reciprocal directions. Fig.\
\ref{fig:SPINS-data} (b) shows constant wave-vector scans measured
along the (0 $K$ 0) direction at 9 K. Very weak dispersion was
observed along this direction and the data show that this excitation
weakens in intensity (significantly) with increasing $K$ and could
not be detected at large $K$. Similar results were obtained along
the ($H$ 1 0) and (0 1 $L$) directions. In order to clarify the
origin of this excitation, polarized neutron scattering experiments
were carried out using the BT7 TAS. As discussed in Ref.
\onlinecite{polarize-technique1,polarize-technique2,polarize-technique3},
coherent nuclear scattering is always non-spin-flip scattering (++)
because it never causes a reversal, or spin flip, of the neutron
spin direction upon scattering. On the other hand, magnetic
scattering depends on the relative orientation of the neutron
polarization $\bf{\hat{p}}$ and the scattering vector $\bf{Q}$. Only
those spin components which are perpendicular to the scattering
vector are effective. Thus for a fully polarized neutron beam, with
the horizontal field configuration, $\bf{\hat{p}}\parallel \bf{Q}$,
all magnetic scattering is spin-flip scattering (-+), and ideally no
non-spin-flip scattering will be observed. Our polarized
measurements were carried out by performing constant wave-vector
scans at (0 1.2 0) at $T$ = 7 K with (++) and (-+) configurations
and a horizontal magnetic guide field at the sample position
($\bf{\hat{p}} \parallel \bf{Q}$). Inset in Fig.\ \ref{fig:polarize}
(a) first compares the (0 2 0) nuclear Bragg scattering measured in
non-spin-flip (++) and spin-flip (-+) configurations. Strong
intensity was observed in (++) channel as expected. The observed
weak (-+) intensity can be attributed to the finite instrumental
flipping ratio which we estimate to be $\sim$ 1/9 by comparing the
integrated intensity of the (-+) and (++) scans of the (0 2 0). The
spin-flip (-+) and non-spin-flip (++) scans at (0 1.2 0) were
plotted in Fig.\ \ref{fig:polarize} (a) between $\hbar\omega$ =
-1.75 meV and 4 meV. Additional magnetic scattering was detected in
the (-+) spin-flip channel. In order to show the peak clearly, the
subtracted data, the non-spin-flip (++) data was subtracted from the
spin-flip (-+) data, is plotted together with the SPINS data in
Fig.\ \ref{fig:polarize} (b). At (0 1.2 0), the SPINS data
(resolution of $\sim$ 0.28 meV) shows an excitation centered at 1.37
$\pm$ 0.05 meV. The subtracted BT7 polarized data (resolution of
$\sim$ 1 meV) shows a rather broad peak consistent with the SPINS
data within experimental error. The polarized measurements indicate
that this low energy excitation is magnetic in origin, which agrees
with the temperature dependence measurements. However, as shown in
Fig. \ref{fig:disp}, it does not fit in the current spin wave model.
Attempting to analyze the combined dispersions with gaps at $\sim$
1.2 meV and $\sim$ 4.7 meV simultaneously using
Eq.~\ref{eq:dispersion} failed, in particular in accounting for the
$\sim$ 1.2 meV excitation. At this time, the nature of the $\sim$
1.2 meV excitation is not clear, and based on it being nearly
dispersionless we can only hypothesize that it may be due to a local
magnetic excitation. Further studies are necessary in order to
unravel the nature of this excitation.

\section{Summary}
The spin dynamics of the ME compound LiCoPO$_4$ were determined by
inelastic neutron scattering experiments. Similar to LiNiPO$_4$
\cite{Jensen2007}, LiMnPO$_4$ \cite{Jiying-Mntobepublish}, and
LiFePO$_4$ \cite{Jiying-PRB-2006}, the overall observed magnetic
excitation spectra in LiCoPO$_4$ can be adequately described by a
linear spin wave theory. Our results indicate that single-ion
anisotropy is as important as the strong nearest-neighbor magnetic
coupling and plays an essential role in understanding the spin
structure and dynamics of LiCoPO$_4$. However, the observation of
the second low energy dispersionless $\sim$ 1.2 meV magnetic
excitation is unusual and is not contained in the spin wave
Hamiltonian, suggesting that it may be closely related to the strong
ME effect in LiCoPO$_4$. The nature of this excitation is not
understood yet and requires further detailed studies.

\section{Acknowledgments}

We acknowledge discussions with T. Barnes. Ames Laboratory is
supported by the U. S. Department of Energy Office of Basic Energy
Science under Contract No. DE-AC02-07CH11358. The HFIR is a national
user facility funded by the United States Department of Energy,
Office of Basic Energy Sciences, Materials Science, under Contract
No. DE-AC05-00OR22725 with UT-Battelle, LLC. SPINS is supported in
part by the National Science Foundation through DMR-0454672. The
work has benefited from the use of the NIST Center of Neutron
Research at the National Institute of Standards Technology.

\end{document}